%% file: main.tex
\documentclass [sigconf]{acmart}
\usepackage{amsmath}
\usepackage{xfrac}
\usepackage{xcolor}
\usepackage{graphicx}
\usepackage{algorithm}
\usepackage{algorithmic}

\AtBeginDocument{%
  }

\setcopyright{acmlicensed}
\copyrightyear{2026}
\acmYear{2026}
\acmDOI{}

\acmConference[Web and Graph Workshop]{WSDM 2026}{WSDM Feb}{2026}
\begin{document}

\title{COFFEE: COdesign Framework for Feature Enriched Embeddings in Ads-Ranking Systems}

\author{Sohini Roychowdhury, Doris Wang, Qian Ge, Joy Mu, Srihari Reddy}
\email{{sroychowdhury1, doriswm, qge2, joymu, sriharir}@meta.com}
\affiliation{\institution{Meta Platforms, Inc.}\country{}}

\input{abstract}

\maketitle


\input{introduction}

\input{methods}
\input{results}

\input{conclusions}

\bibliographystyle{ACM-Reference-Format}
\bibliography{references}

\end{document}

%% file: abstract.tex
\begin{abstract}
Diverse and enriched data sources are essential for commercial ads-recommendation models to accurately assess user interest both before and after engagement with content. While extended user-engagement histories can improve the prediction of user interests, it is equally important to embed activity sequences from multiple sources to ensure freshness of user and ad-representations, following scaling law principles. In this paper, we present a novel three-dimensional framework for enhancing user-ad representations without increasing model inference or serving complexity. The first dimension examines the impact of incorporating diverse event sources; the second considers the benefits of longer user histories; and the third focuses on enriching data with additional event attributes and multi-modal embeddings. We assess the return on investment (ROI) of our source enrichment framework by comparing organic user engagement sources, such as content viewing, with ad-impression sources. The proposed method can boost the area under curve (AUC) and the slope of scaling curves for ad-impression sources by 1.56–2 times compared to organic usage sources even for short online-sequence lengths of $10^{2}$-$10^{4}$. Additionally, click-through rate (CTR) prediction improves by 0.56\% AUC over the baseline production ad-recommendation system when using enriched ad-impression event sources, leading to improved sequence scaling resolutions for longer and offline user-ad representations. 
\end{abstract}

%% file: introduction.tex
\section{Introduction} \label{introduction}
Representation learning, a primary component of modern ads-ranking systems, plays a crucial role in extracting meaningful features from raw data and to accurately gauge user-intent \cite{scalinglaws1, adsformers}. The performance of ads-ranking models is highly correlated to the availability and diversity of high-quality data sources with regards to organic usage, ads-data, and their mutual interactions. Early works on scaling laws from large language models (LLMs) in \cite{scalinglaws1} have demonstrated that the law of resource investment improves linearly by transformer-based representations in the short-term, while improved data encoding and enhanced features lead to long-term exponential improvement on investment. In this work, we define an experimental framework for designing and evaluating multi-modal data sources that are based on user engagement activity and demonstrate the importance of organic and ad-related sources towards enhancement of recommended ad click-through-rates (CTR) \cite{ctr}.

The development of user intent-driven source sequences for large-scale ad-recommendation systems is met with two major challenges. First, sourcing high-quality user-content interaction data with significant global coverage with rapidly evolving multi-modal content sources remains an open challenge \cite{youtube}. Second, the complexity of de-noising multi-modal sources coupled with the rapid evolution of user-behavior and interaction preferences necessitate real-time inferencing \cite{adsformers}. Additionally, the increasing emphasis on data privacy and security has led to stricter regulations and guidelines, further limiting the availability and coverage of certain data sources. In this work, we demonstrate a modular framework to learn personalized user-intention from user engagement activities and to evaluate and boost the learnings by event-level, sequence-level and feature-level enrichments, respectively.

\begin{figure}[ht]
\centering
\includegraphics[width=2.3in, height=2.1in]{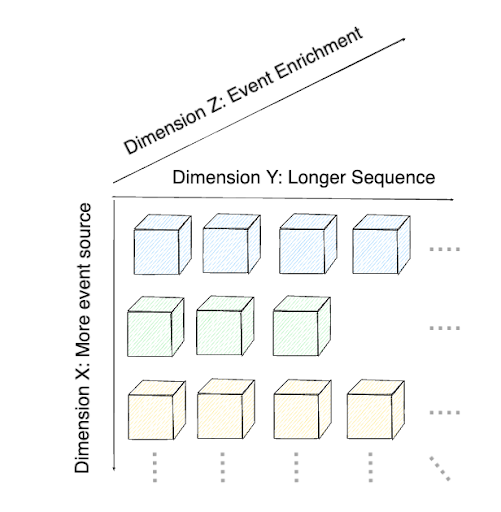}
\caption{Proposed 3-dimensional framework to enrich online-EBF sequences with high adaptability to user-intent.}\label{sys}
\end{figure}

This paper makes two major contributions. First, we present a sequence-learning framework to combine time-stamped data from multi-modal sources gathered from user-engagement activities or events (such as content viewing, likes etc.) into automated event based feature (EBF) sequences. These EBFs enable the sequence-based ads-ranking models to comprehend user-behavior both before and after ad/content interactions for short online sequences \cite{hivq}. Second, we demonstrate 2X improvement in return on investment (ROI) by scaling EBF sequences across dimensions shown in Fig. \ref{sys}.

These dimensions represent scaling user/ad sequences with the following:
\begin{itemize}
\item More events: captures user-intent across multiple personas and multiple surface-level engagements. User-intention profiles can be enriched while maintaining the length of sequences.
\item Longer sequences: captures deeper insights and context regarding user-interests by increasing sequence lengths.
\item Enriched semantics: captures additive nuanced semantic signals for each user-engagement-event and utilizes multi-modal content-embeddings.
\end{itemize}

%% file: methods.tex
\section{Methods and Materials}
In this section, we define EBF sequences from the multi-modal data sources and analyze the impact of the sequences enhancements over deep learning models with pre-trained embeddings in \cite{dlrm}. 

\subsection{The Ads-Recommendation Model}
\begin{figure*}[ht]
\centering
\includegraphics[width=5in, height=4in]{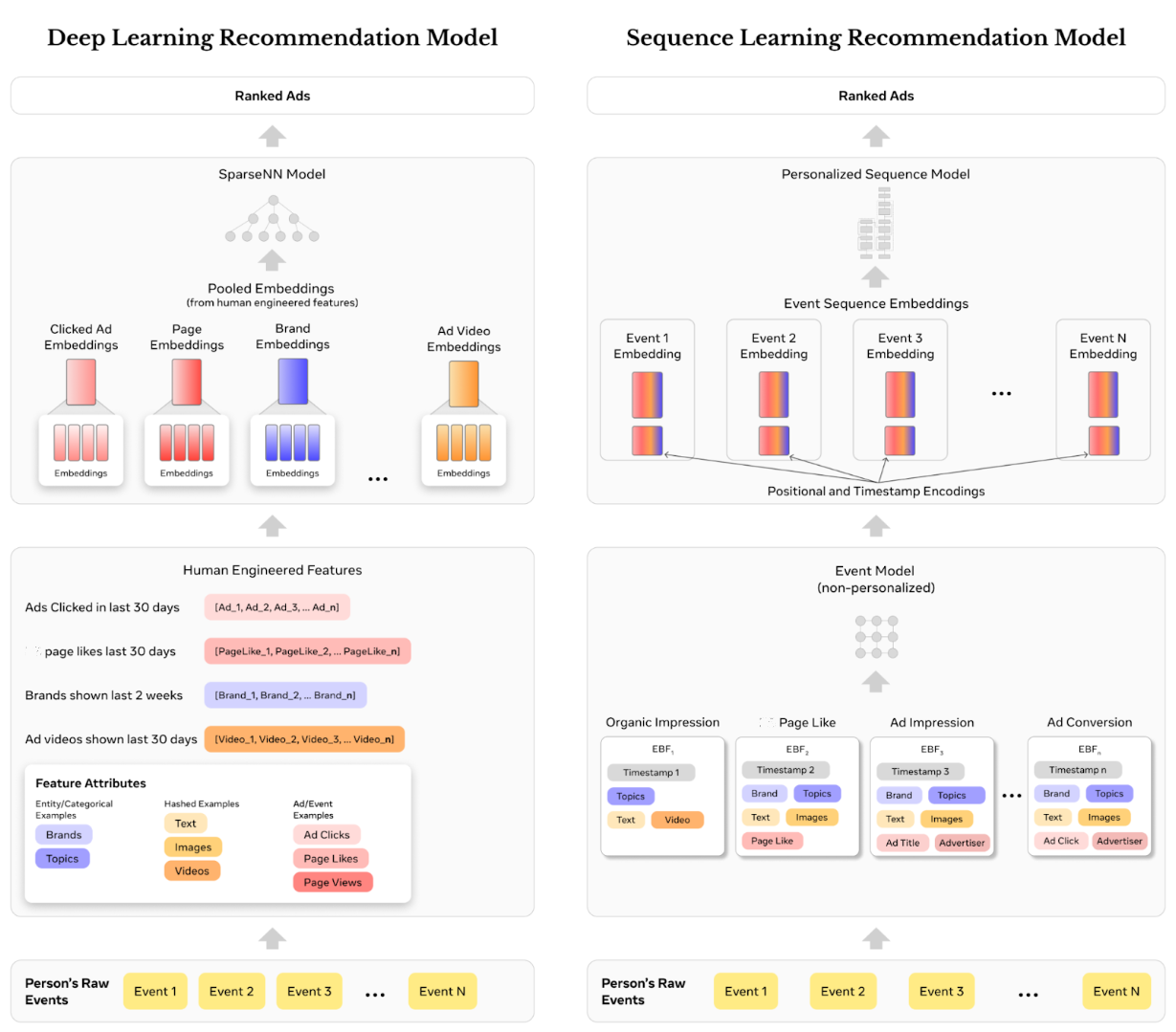}
\caption{Representations for the Sparse NN-based models (left) vs. the Sequence learning Recommendation model (right). }\label{mod}
\end{figure*}

The modified ads-recommendation model used in this work and its structural differences from standard sparse neural network model-based systems \cite{dlrm} are shown in Fig. \ref{mod}. The event-module for the ``\textit{sequence learning recommendation model}'' synthesizes event embeddings from event attributes. So, if an EBF stream has several attributes (topics, brands, advertiser etc. in Fig. \ref{mod}), linear compression is applied to aggregate this information into a single event-attribute based embedding. The event-module then combines the encoded timestamps per user-engagement activity with the single event-attribute based embedding to produce the final event-level representation per-user. Thus, given a list of $r$-timestamp encoded events within a fixed aggregation time window (such as 30 days or 2 weeks in Fig. \ref{mod}), an EBF-event-source of a particular type $n$ (organic impression, likes, ad impression etc. in Fig. \ref{mod}) per user $u$ is defined as a sequence of event-attribute based embeddings across user-activities (with $k$ attributes per event), or $\{a^i_k\}$ in \eqref{ebf}.
\begin{align}\label{ebf}
    EBF^n(u)=[aggregate(\{a_k^i\})], \forall i \in [1,...r], k \in [1,...K].
\end{align}
Here, $i$ represents the $i$-th time-stamped user-engagement activity subjected to attribute level aggregation. To ensure real-time inferences, we limit the number of event attributes to atmost 10. Also, individual events $\{a^i_k\}$ can be raw user-activity features, semantic-ids or content-embeddings as shown in Fig. \ref{mod}.

To further enhance the dynamic nature of user representations, EBF sequence lengths (a combination of attribute length $K$ and event lengths $r$) can be increased. In this work, we extend this method of dynamic user representation from \cite{adsformers} by analyzing the variational impacts of EBF event types ($n$), EBF sequence-lengths ($r$), and EBF-attribute representations ($K$) towards real-time personalization and recommendation at scale. The proposed event-based framework is capable of capturing long-range user-behaviors, such as trends in user-interests and seasonal patterns and scale for longer and offline sequence representations in the order of $10^{6}$ and higher \cite{adsformers}.

\subsection{Sequence Scaling Modules}
We assess the impact of enhancing EBF sequences along the following 3-dimensions towards improved ads/content-recommendation.
 \begin{enumerate}
     \item Scaling with diverse signals by varying $EBF^n$ in \eqref{ebf}: By curating high-value EBFs, the proposed system actively gathers user interaction data from multiple vertical surfaces, such as social media engagement, content consumption, and ad-interactions. This diverse set of signals provides a more holistic view of user intent while capturing both immediate and long-term user preferences as in the prior work \cite{adsformers}.
     \item Scaling with longer sequences by varying $r$ in \eqref{ebf}: Recognizing that longer EBF sequences of user-interactions can provide deeper insights into evolving interests, the proposed system is designed to handle extended sequence lengths. 
     \item Scaling with richer semantics by varying $K$ in \eqref{ebf}: Encoding, concatenating and pooling multi-modal event attributes (text, images, temporal context) into dense embedded sequences results in enriched embeddings. For example, a video-view event is enriched with content embeddings from vision-language models and contextual metadata (e.g., watch duration, device type). Further, customized vector quantization techniques are applied to efficiently encode the attributes of each attribute embedding, thereby resulting in an enhanced and refreshed representations of user-interactions at event-level \cite{hive}.
 \end{enumerate}

\subsection{The EBF Data-sources}\label{sources}
\begin{itemize} 
\item Organic-impression EBFs: This event is generated when user-generated content (UGC) is displayed on the user-screen (website or mobile-app) with $>=50\%$ of the content being visible, and the user views for at least 250 milliseconds.
Event attributes include \{content-id (unique identifier), dwell-time (duration of visibility), media-type (e.g., image, text, video), position (rank in the user’s personalized feed), Timestamp\}. 

\item Ad-impression EBFs: This event is generated when an ad-content is displayed with greater than 50\% visibility and user views for at least 250 milliseconds.
Event attributes include \{semantic-ids (content understanding model-generated metadata for ad enrichment), Ad-id, Timestamp\}

\item Video-view EBF: This event is generated after a user watches a video with $>=30\%$ visibility for at least 1 frame.
Event attributes include \{video-id (unique identifier), author-id, post-id (parent content, if applicable), dwell-time (total view duration), page-id (contextual placement), content-type (ad or UGC), Timestamp\}
\end{itemize}

For EBF enrichment, we curate k-NN of content understanding embeddings as an additional EBF feature \cite{hive}. All EBF experiments below are trained on over 50 billion user-samples curated over a month of usage.

%% file: results.tex
\section{Experiments and Results}
In this section we demonstrate the importance of sequence scaling with EBFs through two sets of experiments. First, we quantify the relative increase in sequence lengths towards CTR improvement vs. increasing training/serving costs. Second, we qualitatively explain the predictive intention improvements for selective user-ad pairs.
\subsection{Quantitative EBF-Source Assessment}
 The ``\textit{sequence learning recommendation model}'' in Fig. \ref{mod} predicts the probability for a user clicking on a piece of content/ad (CTR). We assess the improvement for CTR prediction with varying EBFs in terms of Normalized Entropy (NE), which is defined as:
\begin{equation}
    \textrm{NE} = \frac{\frac{1}{N}\sum \left(y_j \log p_j + (1- y_j) \log (1-p_j)\right)}{\hat{p} \log\hat{p} + \left(1 - \hat{p}\right)\log\left(1 - \hat{p}\right)},
\end{equation}

\noindent where, $N$ is the total number of user-samples, $y_j$ are actual labels, $p_j$ are model predictions and $\hat{p} = \frac{1}{N}{\sum y_j}$ is the prior probability. 

\begin{figure}[ht]
\centering
\includegraphics[width=3in]{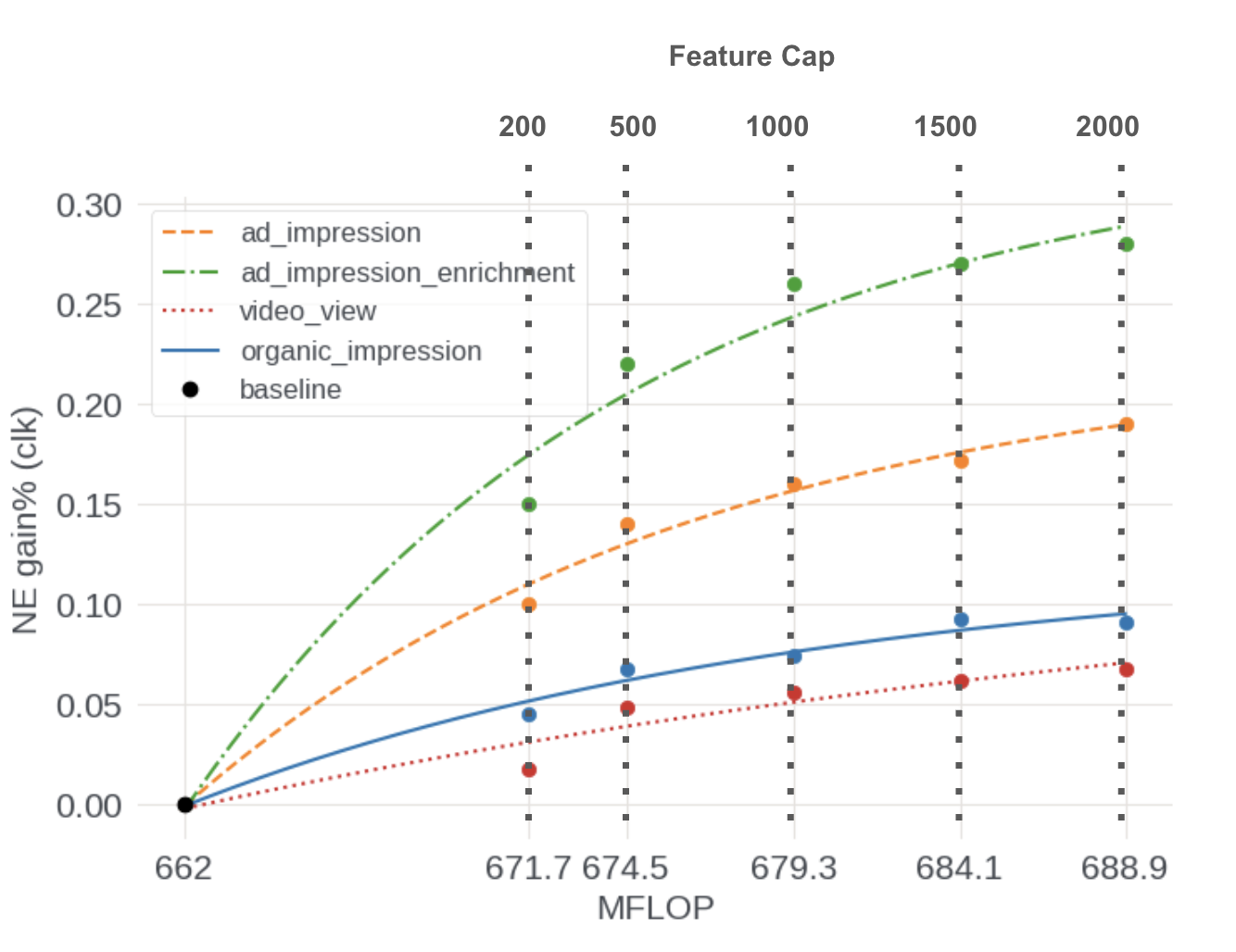}
\caption{Training capacity scaling curves for EBF sources with varying sequence lengths and enrichment-levels. Ad-impression is the strongest source (in green and orange), followed by organic-impression (blue) and video-view (red). The baseline production model performance is depicted by the origin data point in black.}\label{q1}
\end{figure}

We explore the impact of EBF variations for the EBF data sources in Section \ref{sources}. To evaluate the ROI-per EBF source, we utilize the area under the training-capacity scaling-curves (AUC) \cite{scalinglaws1}, as shown in Fig. \ref{q1}, and best fit slope for the training capacity scaling curves (in Table \ref{t1}) as metrics. We evaluate the impact of sequence lengths on the three event sources individually by varying the maximum sequence lengths per EBF in the range of 200 to 2000 (orange, red and blue curves in Fig. \ref{q1}). Also, we evaluate the impact of event enrichment with an additional enrichment feature on top of the existing high-ROI EBF source of ``ad-impression" (green curve in Fig. \ref{q1}). It is noteworthy that similar enrichment-techniques have significantly smaller impact to the video-view and organic-impression EBFs, and are thus omitted from Fig. \ref{q1}. Further, we assess the impact of EBF variations on NE-gains and infrastructure scaling costs through the 3 EBF dimensions: length, nature of events (organic vs. ads) and additional enrichment, respectively.

The following three observations are made across the EBF-variations. First, NE gain curves tend to saturate for longer sequences. Second, the rate of increase in NE gains diminishes with increasing sequence lengths beyond the sequence length of 1000. Third, event-enrichment can further boost NE gains for ad-impressions more than organic-impressions. Additionally, based on the prior work in \cite{adsformers}, we evaluate the AUC and best fit slope of the normalized training capacity scaling curves from Fig. \ref{q1} in Table \ref{t1}.
\begin{center}
\begin{table}[ht]
\caption{ROI for EBFs through AUC and best-fit slopes of training capacity scalng curves.}
\label{t1}
\begin{tabular}{ | c |c | c | } 
\hline
  {\bf EBF Data Sources} & {\bf AUC} &{\bf Slope} \\ 
  \hline
  video view & 0.155 & 1.82 \\ 
  \hline
  organic impression & 0.235 & 2.37 \\ 
  \hline
  ad impression & 0.486 & 4.69 \\ 
  \hline
  ad impression with enrichment & {\bf0.758} & {\bf7.12} \\ 
  \hline
\end{tabular}
\end{table}
\vspace{-0.3in}
\end{center}
Here, we observe that EBF enrichment operations boost AUC and slope of ads-impression source by 1.56x and 1.52x, respectively. Conversely, for organic sources such as video-view and organic-impressions, we observe saturating NE gains as sequence length grows. Also, ROI parameters for the ad-impression source is almost 2x than that of the organic sources. This result is expected since ad-impressions are better indicators of future interest in an ad-per-user when compared to organic sources becoming indicative of future ad-engagement. 

Finally, we analyze the improvement in CTR using the proposed EBF-sequence based model over the baseline \cite{dlrm} method (origin data point in Fig. \ref{q1}). We compare the deployed baseline model with sequence length 200 with the best enriched ad-impression EBF capped to a sequence length of 200. The baseline AUC for CTR prediction is 0.6721, while the best enriched AUC is 0.6759, thereby leading to a 0.56\% increase in CTR prediction AUC at the limited sequence length of 200.
\subsubsection{Limiting condition analysis}
The quantitative analysis demonstrates that both AUC and slope are indicators of ROI per EBF source. Our study extensions involved scaling each source to offline sequences with lengths in the order of 10$^6$ \cite{hivq} to identify the optimal operating conditions per-EBF-source. Also, training longer offline user-ad sequences is 8X times cheaper when compared to shorter online sequences presented here.
\subsection{Qualitative EBF-Source Explainability}
\begin{figure}[ht]
\centering
\includegraphics[width=3in, height=1.1in]{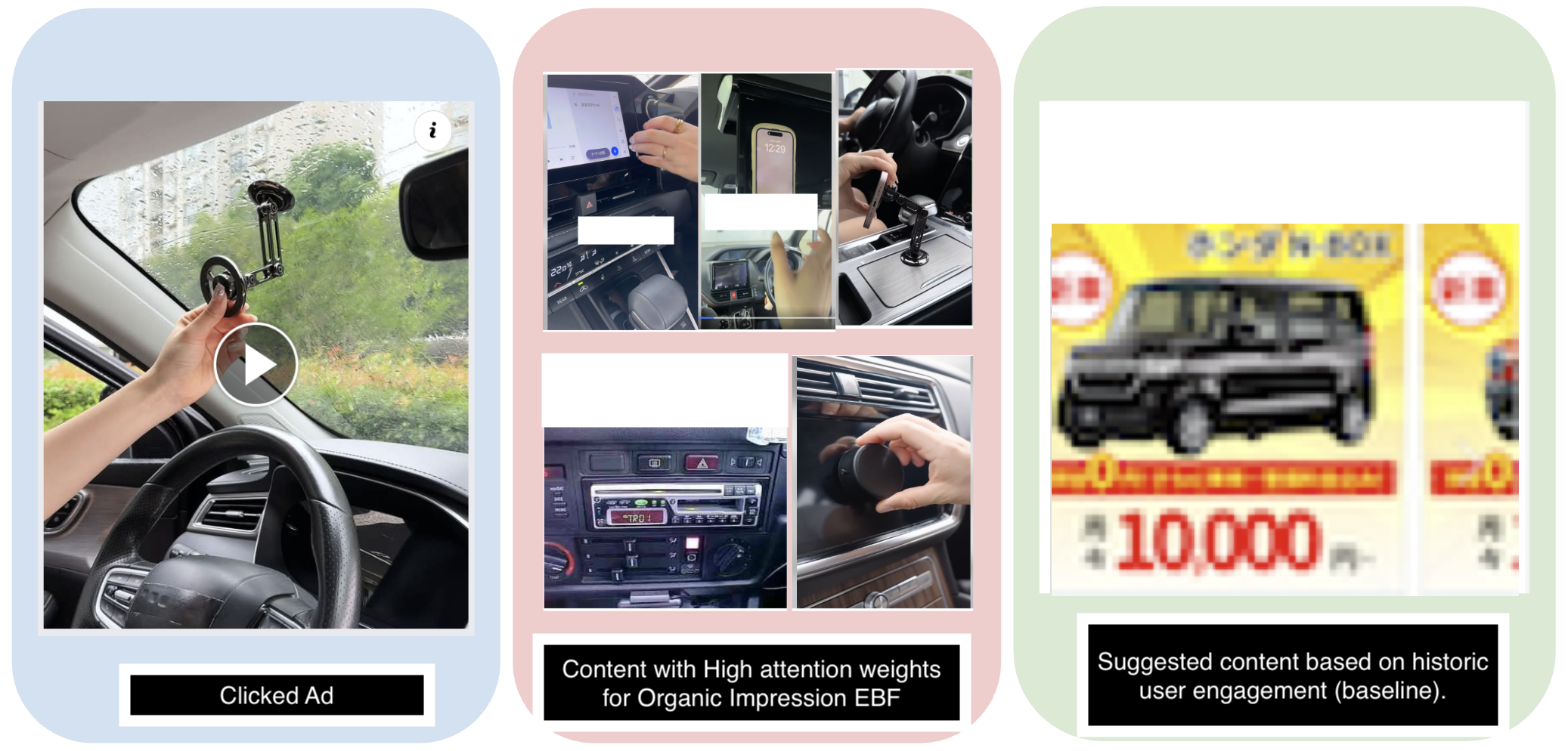}
\caption{Improved ad-prediction using organic impression EBF source vs. historic baseline signals.}\label{ex1}
\vspace{-0.2in}
\end{figure}

\begin{figure}[ht]
\centering
\includegraphics[width=3in, height=1.1in]{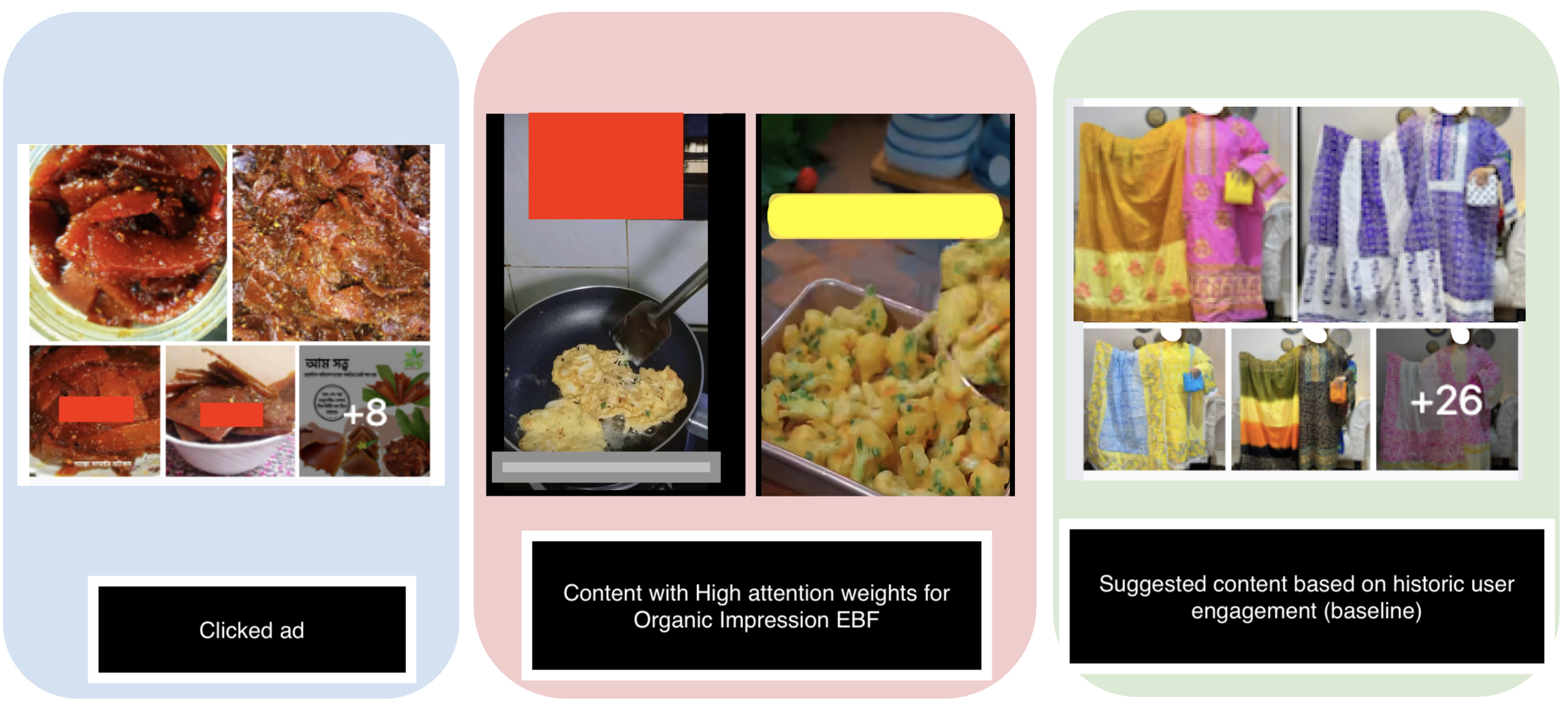}
\caption{User-intent identification for new-ads using organic sequences over baseline.}\label{ex2}
\end{figure}

To assess how EBF sources improve the user-engagement predictions, we utilize cross-attention weights between ads and user-generated sequences. For a specific ad-placement, we extract the historical UGC events with the highest attention weights to assess the relevance between the user history and the served-ad. For the organic impression event source, we analyze the source of NE gains introduced by the new organic-impression EBF-source as opposed to utilizing historic user-engagement activity data-sources. We have the following findings. First, high cross-attention weights capture user-ad intent accurately as shown in Fig. \ref{ex1}. While organic posts with high attention weights (middle image) are related to the served-ad that was clicked by the user (left image), user-ad prediction using historic engagement weights (right image) may have lower dynamic relevance to user intent. This demonstrates the need to personalize EBFs for users rather than reliance on historic ad-engagements.
Second, organic sequences are capable of identifying intent for new ads in the absence of related user-engagement. In Fig. \ref{ex2},  we observe that organic content with high attention scores (middle image) is more-likely to be related to a served-ad that was clicked (left image), when compared to historic engagement-based conversion events (right image).
\vspace{-0.3in}

%% file: conclusions.tex
\section{Conclusions and Discussion}
In this work, we demonstrate that EBF sources (across ad-impressions, organic-impressions and video-views) can be scaled for length, attributes, and event diversity to appropriately model user-and-ad engagement levels that vary widely with time. 
We observe that ad-impression sources with high level representation attributes can improve CTR prediction AUC by 0.56\% at a limited online sequence length of 200, while demonstrating a strong scaling capability with increasing offline sequence lengths to the order of $10^6$ \cite{hivq}. Also, we observe that semantically enriching EBF sequences with content embeddings can further increase the gains from ad-impression EBF sequences and organic sequences. Future works will be directed integrating compressed LLM-based embeddings that can learn from longer and richer user histories for real-time predictions.